\documentstyle[11pt,newpasp,twoside,epsf,epsfig]{article}
\def\ion#1#2{{\rm #1}~{\sc #2}}

\def\edcomment#1{\iffalse\marginpar{\raggedright\sl#1\/}\else\relax\fi}
\reversemarginpar

\begin{document}
\title{Infrared Spectroscopy of Atomic Lines in Gaseous Nebulae
\altaffilmark{1}}
\author{R.~H.~Rubin\altaffilmark{2,3},
~~R.~J.~Dufour\altaffilmark{4},
~~T.~R.\ Geballe\altaffilmark{5},
~~S. W. J. Colgan\altaffilmark{2},\break
J.~P.\ Harrington\altaffilmark{6},
~~S.~D.\ Lord\altaffilmark{7},  
~~A.~L.\ Liao\altaffilmark{2}, and
~~D.~A.\ Levine\altaffilmark{7}}
\altaffiltext{1}{
Based on observations with
ISO, an ESA project with instruments funded by
ESA Member States (especially the PI countries: France, Germany, the
Netherlands and the United Kingdom) with the participation of ISAS and
NASA.  This research was supported in part by ISO Data Analysis funding
from NASA.}
\altaffiltext{2}{NASA/Ames Research Center,
MS 245-6, Moffett Field, CA 94035-1000, USA (rubin@cygnus.arc.nasa.gov)}
\altaffiltext{3}{Orion Enterprises}
\altaffiltext{4}{Rice University}
\altaffiltext{5}{Gemini Observatory}
\altaffiltext{6}{University of Maryland}
\altaffiltext{7}{IPAC, California Institute of Technology}

\begin{abstract}
Spectroscopy in the infrared provides a means to assess important
properties of the plasma in gaseous nebulae.
We present some of our own work that illustrates the need for
interactions between the themes of this conference~-- astronomical
data, atomic data, and plasma simulations.
We undertook Infrared Space Observatory (ISO)
observations with the intent of better understanding the
effects of density variations in nebulae, particularly planetary nebulae
(PNs),
by determining average electron densities from the flux ratios
of several fine-structure, IR emission lines.
Instead, we are able to ascertain only minor density information because
of several instances of the observed line flux ratios being out of range of
the theoretical predictions using current atomic data.
In these cases, the ISO data
cannot presently be used to derive electron density,
but rather provide direction for needed improvements in the atomic
collision strengths.

We have detected an unidentified (uid) strong emission line
in an ISO/SWS spectrum of the Orion Nebula.
The line has a rest wavelength 2.89350$\pm$0.00003~$\mu$m.
A long-slit UKIRT observation confirms the presence of this line
and shows that the emission is spatially extended and appears to be
coincident
with the brightest part of the ionized region.
We do not detect the uid line in our SWS02 spectra of any of the
several bright PNs which we observed for a comparable time.
The need for basic atomic data, in this case
wavelengths to aid species identification, is paramount for future progress.

We look toward the future with a brief synopsis of upcoming or
planned IR missions that promise a significant spectroscopic bearing.
The missions (and their instruments) discussed are SIRTF, SOFIA,
and HERSCHEL (upcoming) and NGST, GSMT, and SAFIR (planned).
\end{abstract}

\section{Introduction}

     Most observational tests of the chemical evolution of the universe
rest on emission line objects; these define the endpoints of stellar
evolution and probe the current state of the interstellar medium.
Gaseous nebulae are laboratories for understanding physical
processes in all emission-line sources, and probes for stellar,
galactic, and primordial nucleosynthesis.

There is a fundamental issue that continues to be problematic~--
the discrepancy between heavy element
abundances inferred from emission lines that are collisionally excited
compared with those
due to recombination/cascading, the so-called ``recombination lines''.
Studies of planetary nebulae (PNs)
contrasting recombination and collisional abundances
(Liu et~al.\ 1995, Kwitter \& Henry 1998)
often find differences exceeding a factor of two.
In an extensive study of NGC~6153, Liu et~al.\ (2000) found
that C$^{++}$/H$^+$,
N$^{++}$/H$^+$,
O$^{++}$/H$^+$, and
Ne$^{++}$/H$^+$ ratios derived from optical recombination lines
are all a factor of $\sim$10 higher than the corresponding values
deduced from collisionally-excited lines.
Abundances determined from these two methods
disagree by a factor larger than
the spread of abundances used to determine such fundamental
quantities as Galactic abundance gradients (e.g., Shaver et~al.\ 1983;
Simpson
et~al.\ 1995; Henry \& Worthey 1999).

Most of the efforts to explain the abundance
puzzle between collisional and recombination values have attempted
to do so by examining electron temperature ($T_e$) variations in the plasma.
This is often done, using the formalism of Peimbert (1967), in terms
of the mean-square variation ($t^2$) of $T_e$.
The inferred metallicity obtained by using the usual (optical/UV)
forbidden lines is very sensitive to $T_e$ (exponential) and $t^2$.
On the other hand, recombination lines
are rather insensitive to $T_e$ and $t^2$.
Agreement  close to the higher recombination value
can be forced in the derived abundance
by attributing the difference to (solving for) $t^2$.
For instance, consider the case of the PN NGC~7009, which
has stood near the center of the abundances controversy.
Liu et~al.\ (1995) found
that the recombination C, N, and O abundances are a factor of $\sim$5
larger than the corresponding collisional abundances.
By invoking
$t^2$~$\sim$ 0.1 for oxygen in NGC~7009,
agreement can be forced in the derived abundance
close to the higher recombination value -- a
value more than 2.5 times larger than the solar
O/H of 7.41$\times$10$^{-4}$ (Grevesse \& Sauval 1998).
Such a large $t^2$ is not at all predicted by current theory/models
(e.g., Kingdon \& Ferland 1998).

For NGC~7009, Rubin et~al.\ (2000) produced a $T_e$ map
on a pixel scale of 0.1$''$
from the flux ratio of [\ion{O}{iii}] 4363/5007 using a set
of HST/WFPC2 images that included narrow-band filters
F437N and F502N to assess these line fluxes.
Their analysis of $t^2$ in the plane of the sky
indicates very small values,
{\bf $\la$ 0.01},
throughout the nebula.
It seems clear that something more than $T_e$ variations alone is necessary
to explain the abundance dilemma.

We have been addressing other ingredients
that may affect the determination of elemental abundances from observations.
Perhaps the most important of these is density variations.
While $T_e$ variations are expected theoretically to be fairly small
in PNs and
\ion{H}{ii}
regions, variations in electron density ($N_e$)
are almost certainly very large.
High $N_e$ values ($\sim$10$^6$~cm$^{-3}$) are inferred
in the cometary knots of the Helix nebula
(e.g., Walsh \& Meaburn 1993; Burkert \& O'Dell 1998)
and around proplyds in Orion
(e.g., Henney \& O'Dell 1999 and references therein).
Gas at $N_e$~= 10$^6$~cm$^{-3}$ will produce
10$^6$ times as much emission as the same volume of
``normal" nebular gas
at $N_e$~= 10$^3$~cm$^{-3}$
for  recombination lines and collisionally-excited lines
that are not suffering collisional
deexcitation~-- when $N_e$~$<$ $N_{crit}$ (the critical density).

Several well known line ratios of collisionally-excited lines
serve as diagnostics of $N_e$,
probing different ionization conditions and different density regimes.
Because various collisionally-excited lines undergo
collisional deexcitation at different $N_{crit}$ values,
the derivation of an average $N_e$ for a given line ratio
depends
on the specific $N_e$ (and $T_e$) dependence of the respective volume
emissivities.
In general, different average $N_e$ values are obtained for various line
pairs due to different contributions to the observed
intensities from the volume observed.
Very substantial biases
in the inference of abundances from collisionally-excited lines may then
result
(Rubin 1989; Viegas \& Clegg 1994; Liu et~al.\ 2000).


\begin{figure}
\plotfiddle{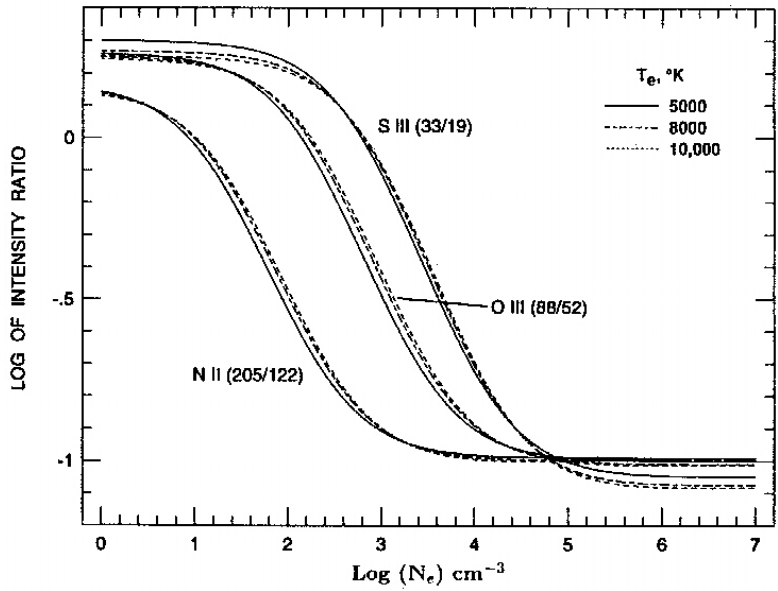}{3.0truein}{0}{40}{40}{-130}{-40}
\vskip-0.1truein
\caption{Three valuable indicators of $N_e$ --
the intensity ratio of [\ion{N}{ii}] (205/122), [\ion{O}{iii}]
(88/52), and [\ion{S}{iii}] (33/19) vs.\ $N_e$~(cm$^{-3}$).
The [\ion{N}{ii}] ratio is most sensitive
at the lowest $N_e$, and [\ion{S}{iii}] at the highest.}
\vskip-0.1truein
\end{figure}

\section{The Promise of the Infrared for Density Diagnostics}
Measurement of the
two $\Delta$J~=~1
IR fine-structure transitions for species with a $^3P$ ground
state provides a diagnostic for $N_e$.
The ratio of their fluxes is a sensitive indicator of $N_e$ (over some range
in
$N_e$) that generally is not significantly affected by $T_e$ or extinction.
For example, the flux ratios
[\ion{O}{iii}] 52/88~$\mu$m,
[\ion{S}{iii}] 19/33~$\mu$m,
and
[\ion{N}{ii}]  122/205~$\mu$m
provide $N_e$[\ion{O}{iii}],
$N_e$[\ion{S}{iii}],
and $N_e$[\ion{N}{ii}], respectively, as shown in Figure~1 (Rubin et~al.\
1994).

There is only a weak dependence on $T_e$ as illustrated for each species
by the three curves for $T_e$~= 5000, 8000, and 10000~K.
For conditions prevalent in many
PNs and \ion{H}{ii} regions, these lines
readily show the effect of collisional deexcitation because
some of the plasma may exceed
their respective  $N_{crit}$-values.
We are able to use only two of the above diagnostics because ISO did not
observe the 205~$\mu$m line.


\begin{figure}
\plotfiddle{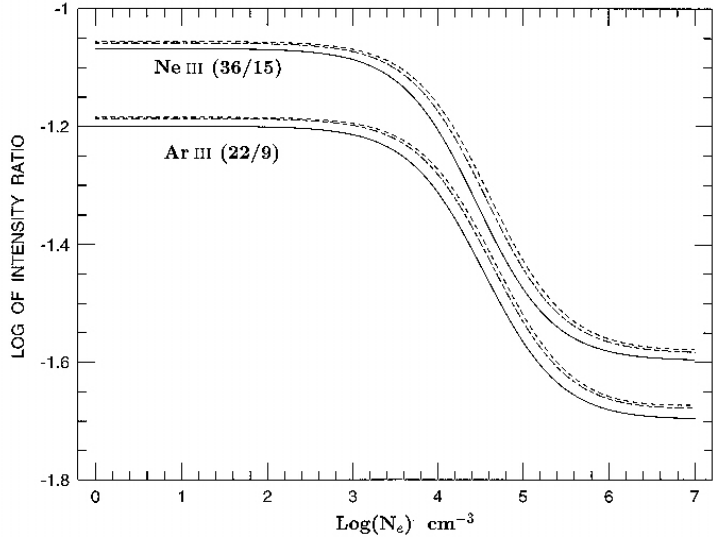}{3.0truein}{0}{40}{40}{-130}{-30}
\vskip-0.3truein
\vspace{-6pt}
\caption{Similar to Fig.~1 for two other
indicators of $N_e$~-- the intensity ratio of [\ion{Ne}{iii}]
(36/15) and [\ion{Ar}{iii}] (22/9).  These diagnostics are
now available from ISO.  These ratios are sensitive indicators of
higher density material than those ratios in Fig.~1.}
\vskip-0.2truein
\end{figure}

With ISO, there are now at least three important additional IR line
ratios available~--
[\ion{Ne}{iii}] 36.0/15.5, [\ion{Ne}{v}] 14.3/24.3,
and [\ion{Ar}{iii}] 21.8/8.99~$\mu$m~--
for studies of $N_e$ structure and variations.
This is because ISO was able to observe the [\ion{Ne}{iii}] 15.5~$\mu$m,
[\ion{Ne}{v}] 14.3~$\mu$m, and [\ion{Ar}{iii}] 21.8~$\mu$m lines, which
suffer very seriously from the atmosphere even for airborne astronomy,
e.g., for the Kuiper Airborne Observatory (KAO) or SOFIA.
(Observations of these lines with the
IRAS LRS were sparse and of dubious quality.)
These three ratios are sensitive indicators
of higher density material than are the other three ratios mentioned
earlier.
Two of these ratios -- [\ion{Ne}{iii}] 36/15~$\mu$m and [\ion{Ar}{iii}]
22/9~$\mu$m --
are displayed in Figure~2.

The ratios are similar in that they
are most discriminant in the range
3.8 $\la$ log~$N_e$
$\la$ 5.4.
We will use two more $N_e$-diagnostic line ratios
[\ion{Mg}{v}] 13.5/5.61~$\mu$m and [\ion{Ar}{v}] 13.1/7.9~$\mu$m.
These will be used to probe the higher ionization gas.
Although usually weaker than the lines mentioned earlier,
all are expected to be seen in a large number of high ionization PNs.
The hope for  the ISO IR data is that these numerous ``new" $N_e$
diagnostics,
in combination with the $N_e$ diagnostics
available from the optical/UV,
will permit a
viable tomographic
analysis of the $N_e$ structure in PNs.

\section{ISO Observations and Results for the Density Analysis}
Under our GO programs, we observed
the PNs NGC~2022, NGC~6210, NGC~6818,
and IC~2165
with the ISO Short Wavelength  Spectrometer (SWS).
Our observations with SWS were made in SWS02 mode, which provides higher
spectral resolution than SWS01 mode.
One of our goals was to derive $N_e$ from several
diagnostic line pairs in order to address density variations within
these objects.  As described above, we aimed to obtain
$N_e$ values from the following six flux ratios:
[\ion{S}{iii}]  (18.7/33.5),
[\ion{Ar}{iii}] (8.99/21.8),
[\ion{Ne}{iii}]  (15.5/36.0),
[\ion{Ar}{v}]  (7.90/13.1),
[\ion{Ne}{v}] (14.3/24.3),
and [\ion{Mg}{v}] (5.61/13.5$\mu$m).
Because the ISO/SWS aperture size
depends on wavelength, the above line
fluxes are not always directly comparable.
For the lines discussed here, the sizes were:
14$''$~$\times$~20$''$ for the 5.61, 7.90, \& 8.99~$\mu$m lines;
14$''$~$\times$~27$''$ for the 13.1~-- 24.3~$\mu$m  lines;
and
20$''$ $\times$ 33$''$  for the 33.5 \& 36.0~$\mu$m lines.
Table~1 provides a summary of which lines were observed in the various
sources.


\begin{table}
    \caption{~~~~~}
\plotfiddle{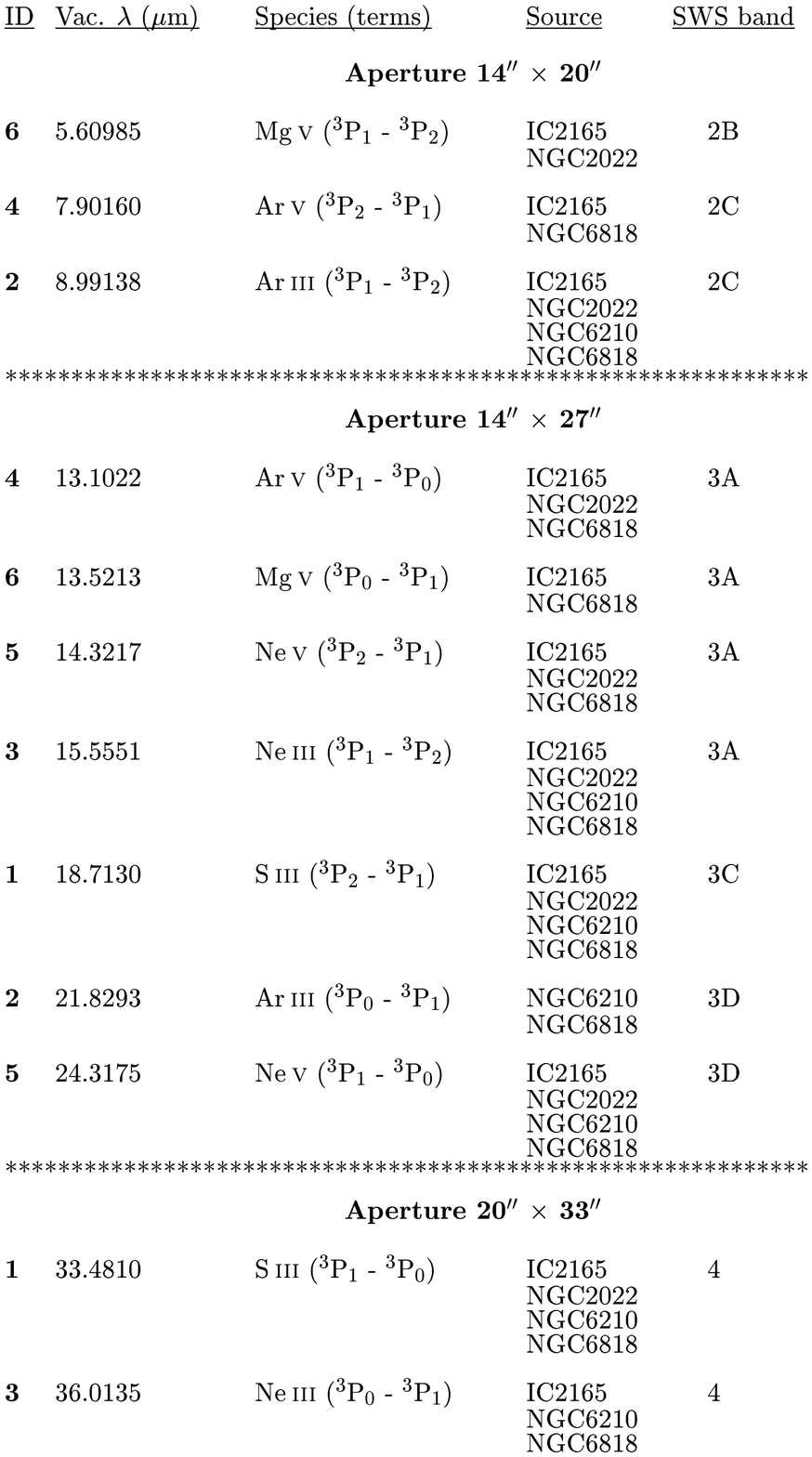}{4.0truein}{0}{40}{40}{-130}{0}
\vskip-0.3truein
\end{table}

\vskip0.3truein

The position angles (PA) of the long-axis of the various apertures
were
183.17,
13.27,
349.07,
and
193.71$^{\rm o}$
for
NGC~2022, NGC~6210, NGC~6818, and IC~2165, respectively,
and rough estimates of the respective source ``diameters" are
19$''$, 16.2$''$, 20$''$, and 9$''$.
However, there is certainly emission well beyond these ``diameters"
in the case of NGC~6210~-- see an overlay\break
(http://www-space.arc.nasa.gov/$\sim$rubin/turtle.html)
of our ISO apertures on HST/WFPC2 images
taken by us under GO-6792~-- and NGC~6818~--\break
(http://www-space.arc.nasa.gov/$\sim$rubin/zorro.html)
other WFPC2 images under GO-6792.
The best object for our analysis in this sense is IC~2165,
which is smaller than the smallest aperture.
For the other PNs, the use of line flux ratios may still be valid as
long as the specific ionic emitting zone is circumscribed by the
smaller of the two apertures.
Of the six line sets, {\it only}
the [\ion{Ne}{v}] pair of lines (14.3, 24.3) was
observed with the same aperture size.


\begin{figure}
\plotfiddle{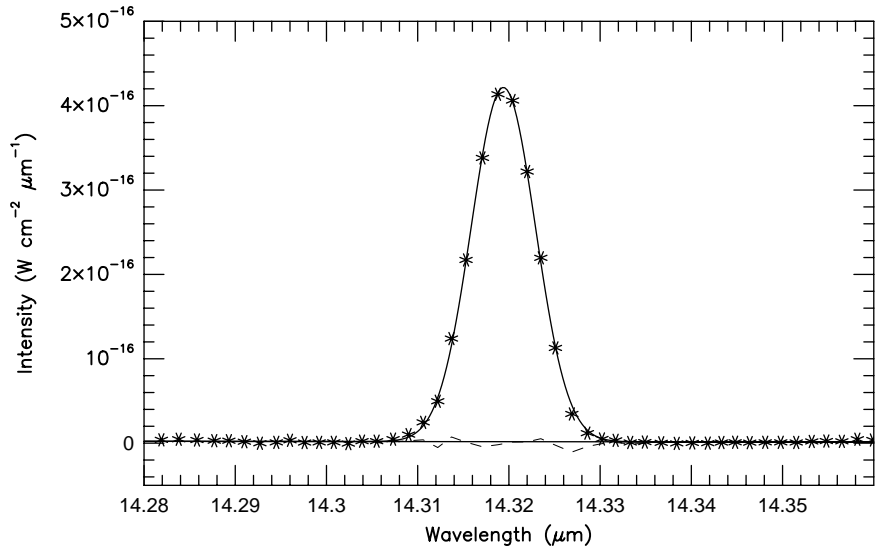}{3.0truein}{0}{50}{50}{-190}{-25}
\vskip-0.5truein
\caption{ISO SWS02 spectrum of [\ion{Ne}{v}]
14.32~$\mu$m line in the PN NGC~6818.
This important $N_e$-diagnostic line cannot be observed
even from an airborne platform.
The data (asterisks) have been fit with a Gaussian and a linear baseline.
The residuals, data point minus curve fit, are indicated as the dashed
line.}
\vskip-0.2truein
\end{figure}

Figures 3 and 4 are ISO SWS02 spectra of the [\ion{Ne}{v}] 14.32
and 24.32~$\mu$m lines in NGC~6818.
The aperture used for each was 14$''$ $\times$ 27$''$.
The data are fit very well by a Gaussian profile.
Data processing was performed using the ISO Spectral Analysis Package
(ISAP).
NGC~6818 is not fully enclosed by a 14$''$ $\times$ 27$''$ aperture.
However, the same portion of the nebula is observed in both of these lines.
Because the high ionization Ne$^{+4}$ zone is much more
concentrated toward the central star than the
entire ionized (\ion{H}{ii}) region is, it is
highly probable that the total Ne$^{+4}$
zone is enclosed within the observed aperture.

Figure~5 shows  the theoretical flux ratio F(14.3)/F(24.3)
vs.\ $N_e$~(cm$^{-3}$), using
effective collision strengths from Lennon \& Burke (1994).
Throughout this paper, we shorten
effective collision strength ($\Upsilon$) to collision strength (CS).
Our ISO data with regard to this ionic species are most interesting.
In the subsections to follow, we present the SWS data and discuss the
six ionic species in order of lowest to highest degree of ionization
as indicated by the range in ionization potential (I.P.) over which the
species exist.

\subsection{\ion{S}{iii}}

Observations were made of both the
[\ion{S}{iii}] 18.7 and 33.5~$\mu$m lines in all four of our program PNs.
The line profiles were fit with the line-fitting routine in ISAP,

\begin{figure}
\plotfiddle{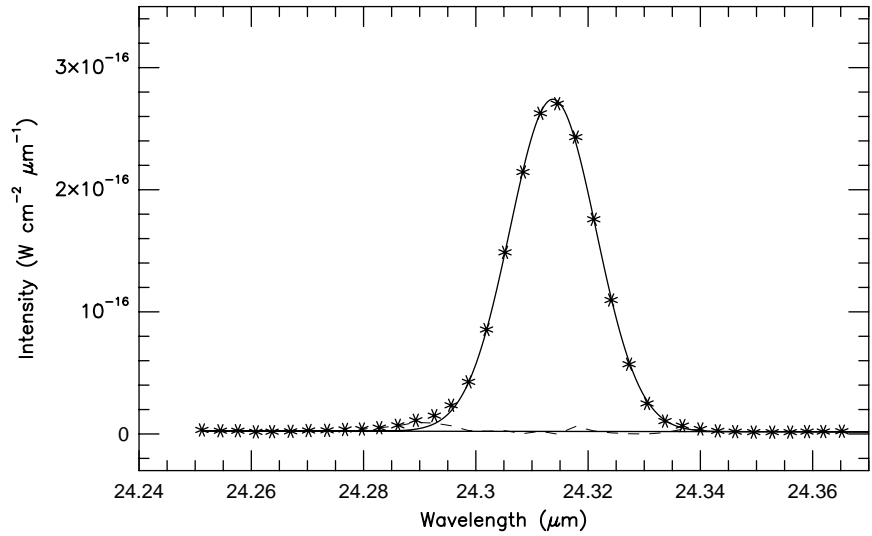}{225pt}{0}{50}{50}{-190}{-55}
\caption{Similar to Fig.~3 for the ISO SWS02 spectrum of [\ion{Ne}{v}]
24.32~$\mu$m line in the PN NGC~6818.}
\end{figure}
\begin{figure}
\plotfiddle{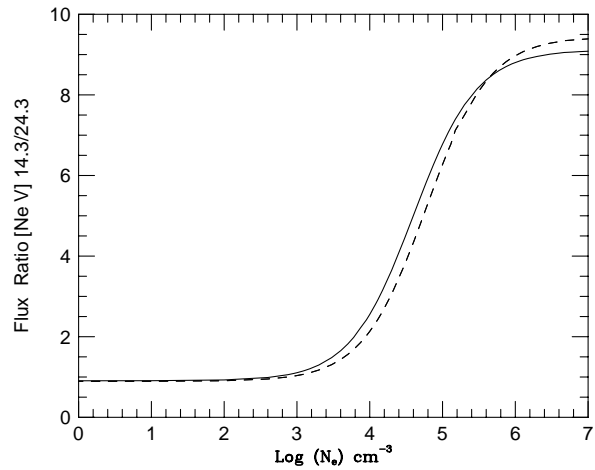}{225pt}{0}{40}{40}{-120}{-110}
\caption{An outstanding indicator of $N_e$ is the flux ratio of
[\ion{Ne}{v}]
(14.3/24.3).
This shows  the theoretical flux ratio F(14.3)/F(24.3) vs.\
$N_e$~(cm$^{-3}$)
using Lennon \& Burke (1994) collision strengths.
The ratio is insensitive to $T_e$ (10,000~K solid, 15,000~K dashed curve).
This diagnostic is now available from ISO data.}
\end{figure}
\noindent
which provides the line flux and uncertainty.
We fit the data points with a Gaussian and a linear baseline such as
shown in Figs.\ 3 and 4 for the [\ion{Ne}{v}] data.
In general, the data were fit well by this method.
Table~2 presents the measured fluxes and uncertainties.
These uncertainties do not include systematic errors.
The last column has the $N_e$ derived from the flux ratio assuming
a $T_e$ of 10,000~K.
It is likely that only the $N_e$ for IC~2165 is valid.
Because of the extended [\ion{S}{iii}] emission and
the mismatch of aperture sizes, the F(18.7)/F(33.5) ratio
is likely a lower limit and the $N_e$ entry a lower limit
for the other objects.

In the lower part of Table~2, we provide the theoretical
F(18.7)/F(33.5) ratio in the low-$N_e$ limit; there, the smallest value
of the ratio obtains.  The transitions are shown in a term diagram
(not to scale) with $N_{crit}$-values for levels 2 and 3
given for $T_e$~= 10,000~K.


\vskip0.5truein
\begin{table}
    \caption{~~~~~}
\plotfiddle{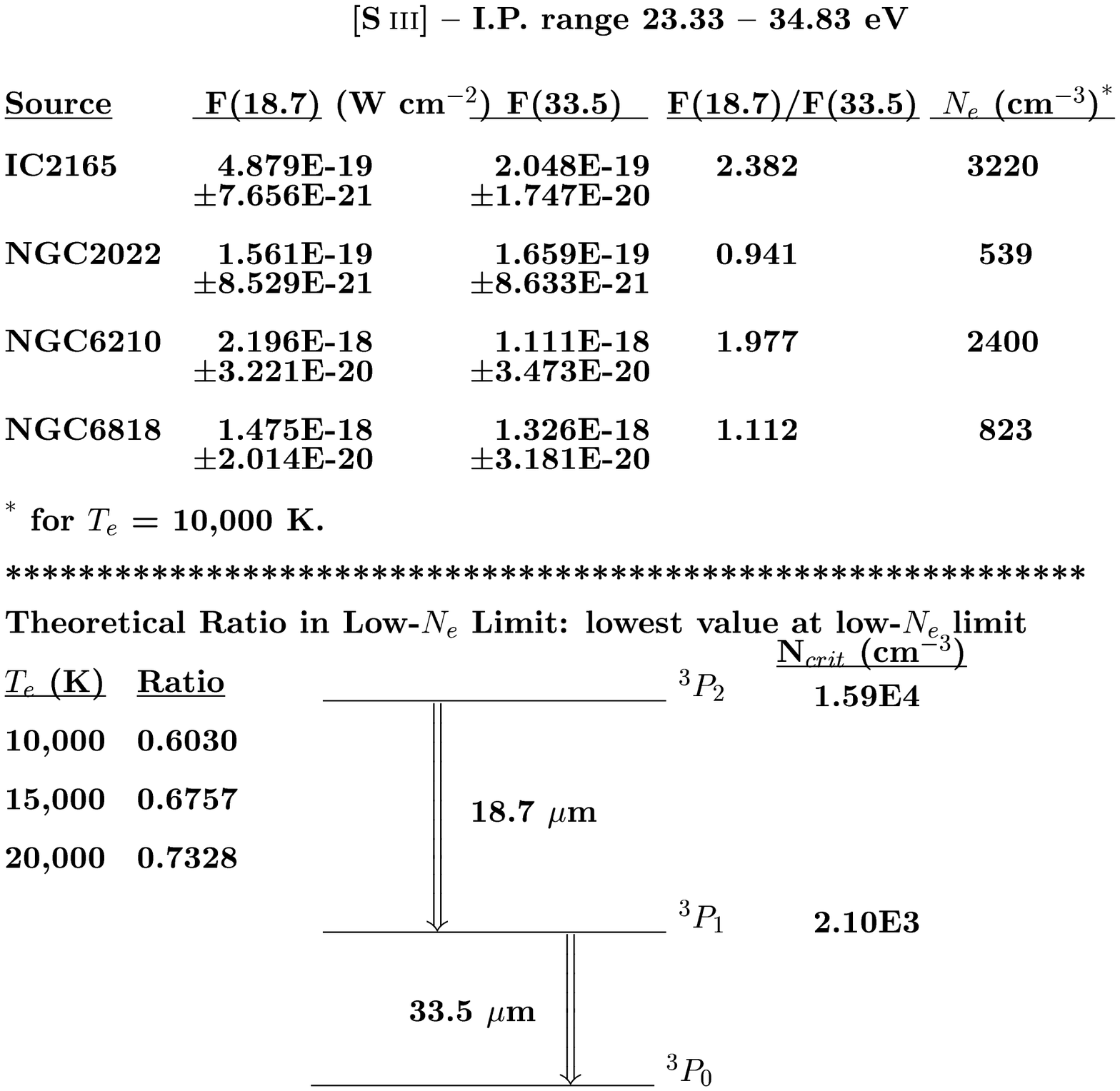}{3.0truein}{0}{40}{40}{-120}{0}
\vskip-1.truein
\end{table}

\subsection{\ion{Ar}{iii}}

Observations were made of the
[\ion{Ar}{iii}] 8.99 and 21.8~$\mu$m lines in NGC~6210 and NGC~6818.
Table~3 presents the measured fluxes for these data.
The comments regarding line fitting,  etc., and the tabular entries
that accompanied Table~2 apply here and for the ionic species to follow.
In the lower part of the Table, we have the theoretical
F(21.8)/F(8.99) ratio in the low-$N_e$ limit, where the highest value
of the ratio obtains.
      For NGC~6210, the observed ratio is close to the low-$N_e$
asymptotic limit, where the ratio is insensitive to $N_e$ and thus not
useful for deriving $N_e$.  For this and other such instances,
``near low limit" is entered in the $N_e$-column of the Table.
For NGC~6818, the observed ratio exceeds the low-$N_e$ limit
for any reasonable $T_e$.
Because the [\ion{Ar}{iii}] emission is likely more extended than the
smaller 14$''$ $\times$ 20$''$ aperture used for the 8.99~$\mu$m line, the
F(21.8)/F(8.99) ratio is likely an upper limit
and thus any inference
that the observed ratio is {\it out of the theoretical bounds}
is uncertain.  We enter ``XXX~low~?" in the Table.


\begin{table}
    \caption{~~~~~}
\plotfiddle{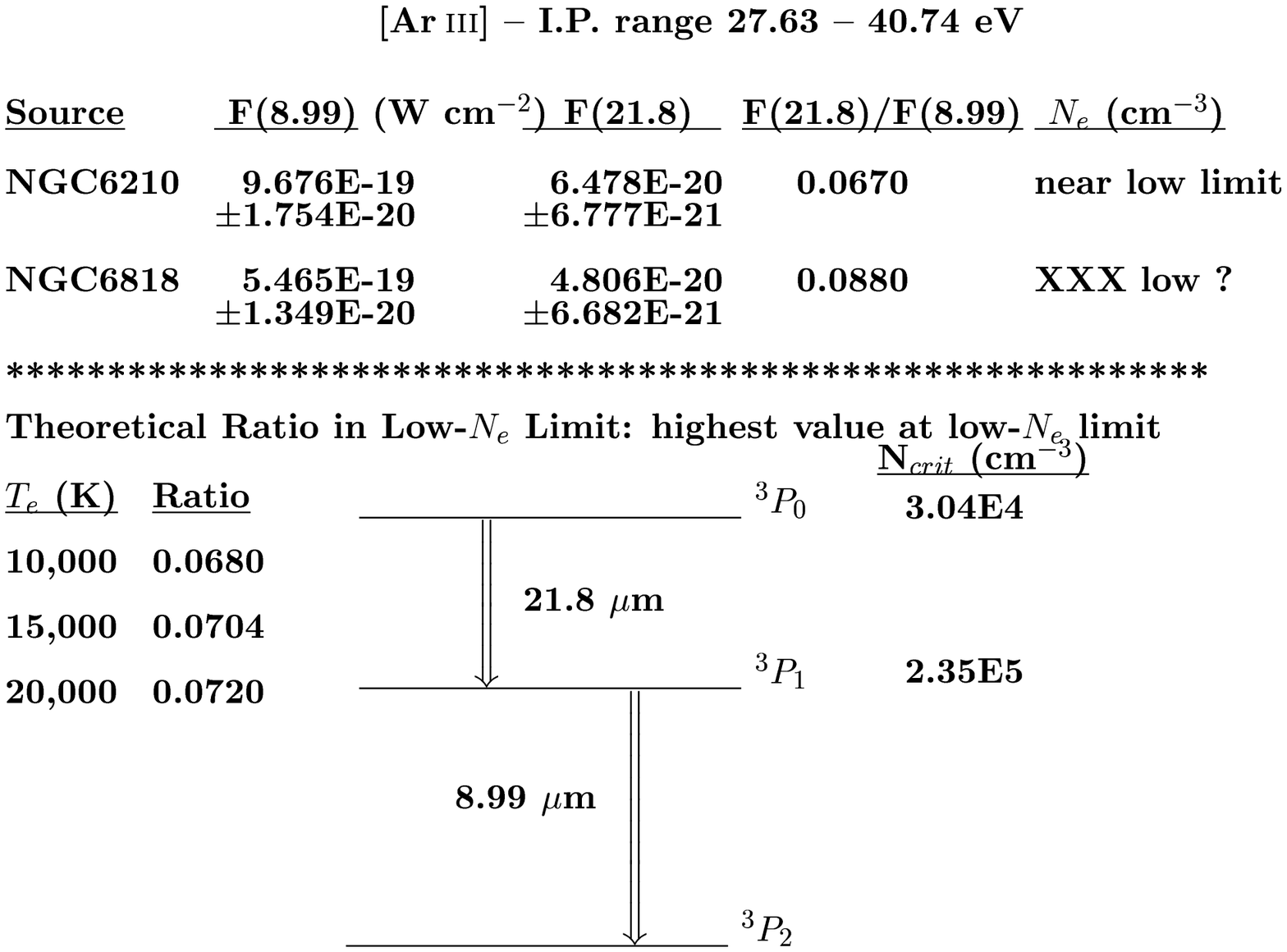}{3.0truein}{0}{40}{40}{-120}{-40}
\vskip-1.0truein
\end{table}

\subsection{\ion{Ne}{iii}}

Observations were made of the
[\ion{Ne}{iii}] 15.5 and 36.0~$\mu$m lines in IC~2165, NGC~6210, and
NGC~6818.
Table~4 presents the measured fluxes for these data
as well as the theoretical
F(36.0)/F(15.5) ratio in the low-$N_e$ limit, where the highest value
of the ratio obtains.
For all the sources, the observed ratio exceeds the low-$N_e$ limit.
However because the [\ion{Ne}{iii}] emission
for NGC~6210 and NGC~6818
may be more extended than the
smaller 14$''$ $\times$ 27$''$ aperture used for the 15.5~$\mu$m line,
the F(36.0)/F(15.5) ratio is likely an upper limit and
again has an uncertain (XXX~low~?) conclusion.
On the other hand, for IC~2165, we are dealing with integrated fluxes
and conclude robustly
that the observed ratio is {\it out of the theoretical bounds}
and enter ``XXX low" in the Table.
The theoretical ratio in the low-$N_e$ limit is set by the effective
collision strengths.
We suggest the data for IC~2165 are pointing to a need to reexamine the CS,
particularly those between the $^3P$ fine-structure levels.


\begin{table}
    \caption{~~~~~}
\plotfiddle{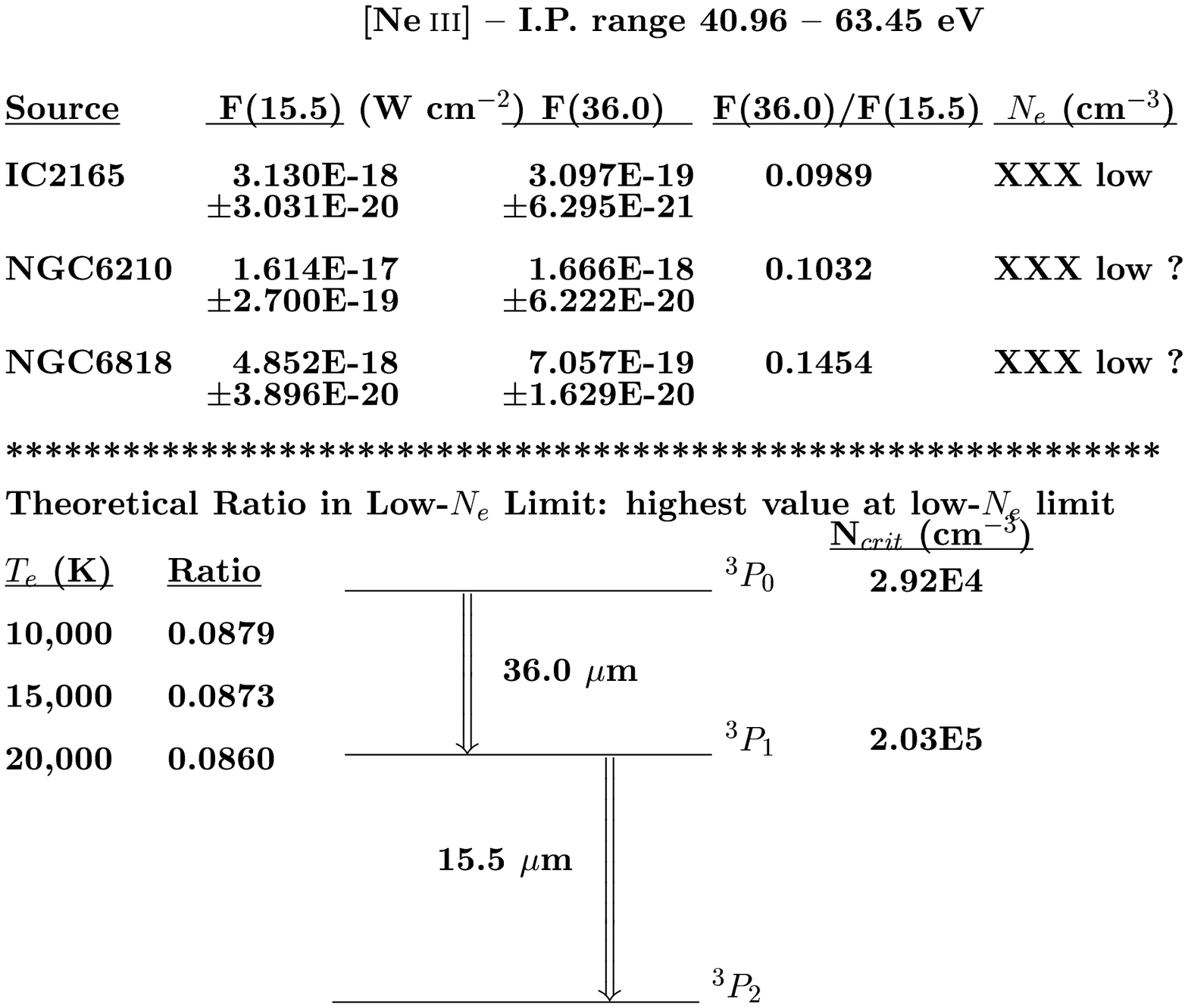}{3.0truein}{0}{40}{40}{-120}{-40}
\vskip-0.7truein
\end{table}

\vskip-0.6truein

\subsection{\ion{Ar}{v}}

Observations were made of the
[\ion{Ar}{v}] 7.90 and 13.1~$\mu$m lines in IC~2165 and NGC~6818.
Table~5 presents the measured fluxes for these data as well as the
theoretical
F(7.90)/F(13.1) ratio in the low-$N_e$ limit, where the lowest value
of the ratio obtains.
For both PNs, the observed ratio is near the low-$N_e$ limit.
Here, even in the case of NGC~6818,
it is likely that the Ar$^{+4}$ zone is much more centrally concentrated
than the size of the H$^+$ zone;
but if the [\ion{Ar}{v}] 7.90 emission
were more extended than the
smaller 14$''$ $\times$ 20$''$ used,
then the F(7.90)/F(13.1) ratio would be a lower limit.

\vspace{-5pt}
\begin{table}
    \caption{~~~~~}
\plotfiddle{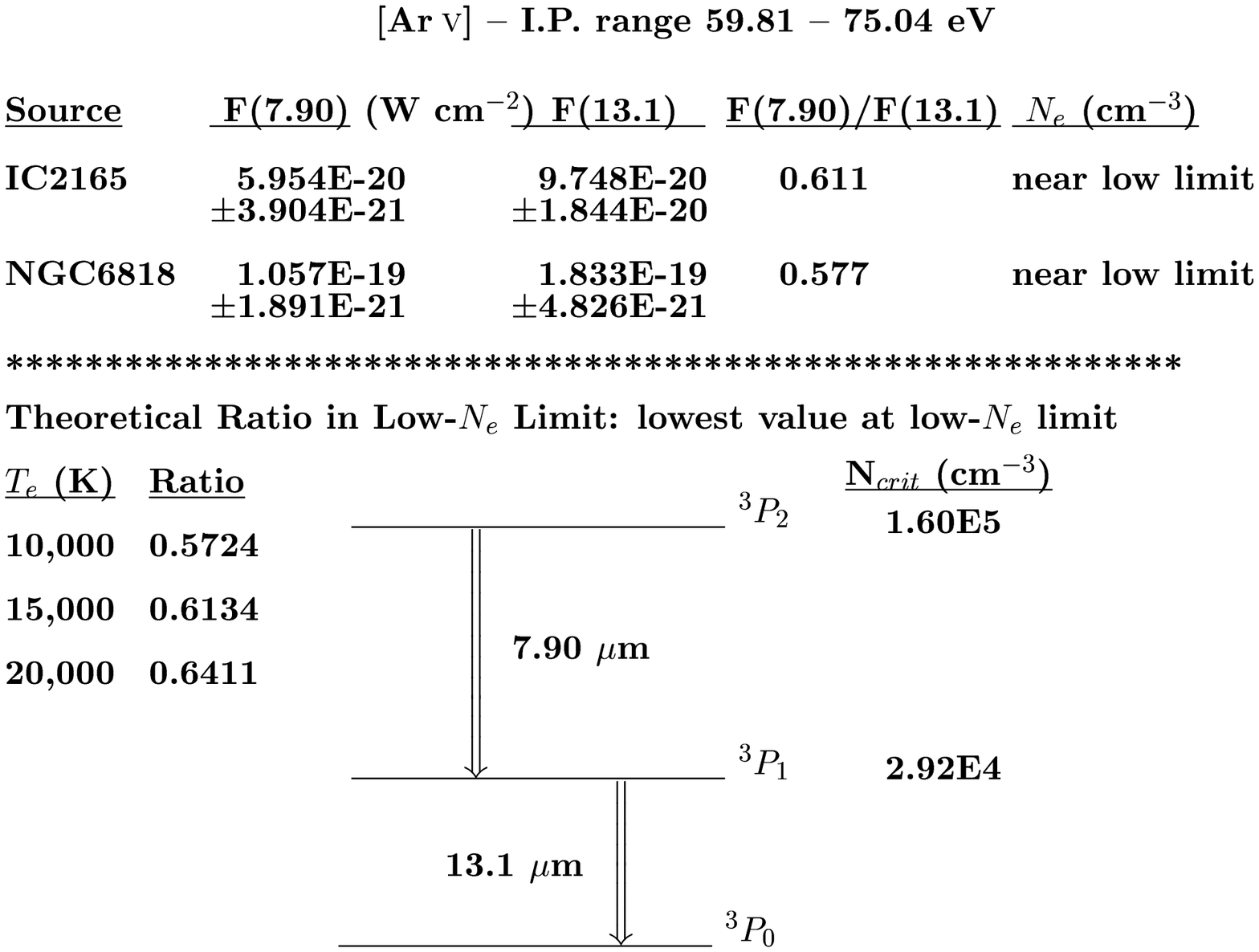}{3.0truein}{0}{40}{40}{-120}{-40}
\vskip-0.6truein
\end{table}
\vspace{-10pt}
\begin{table}
    \caption{~~~~~}
\plotfiddle{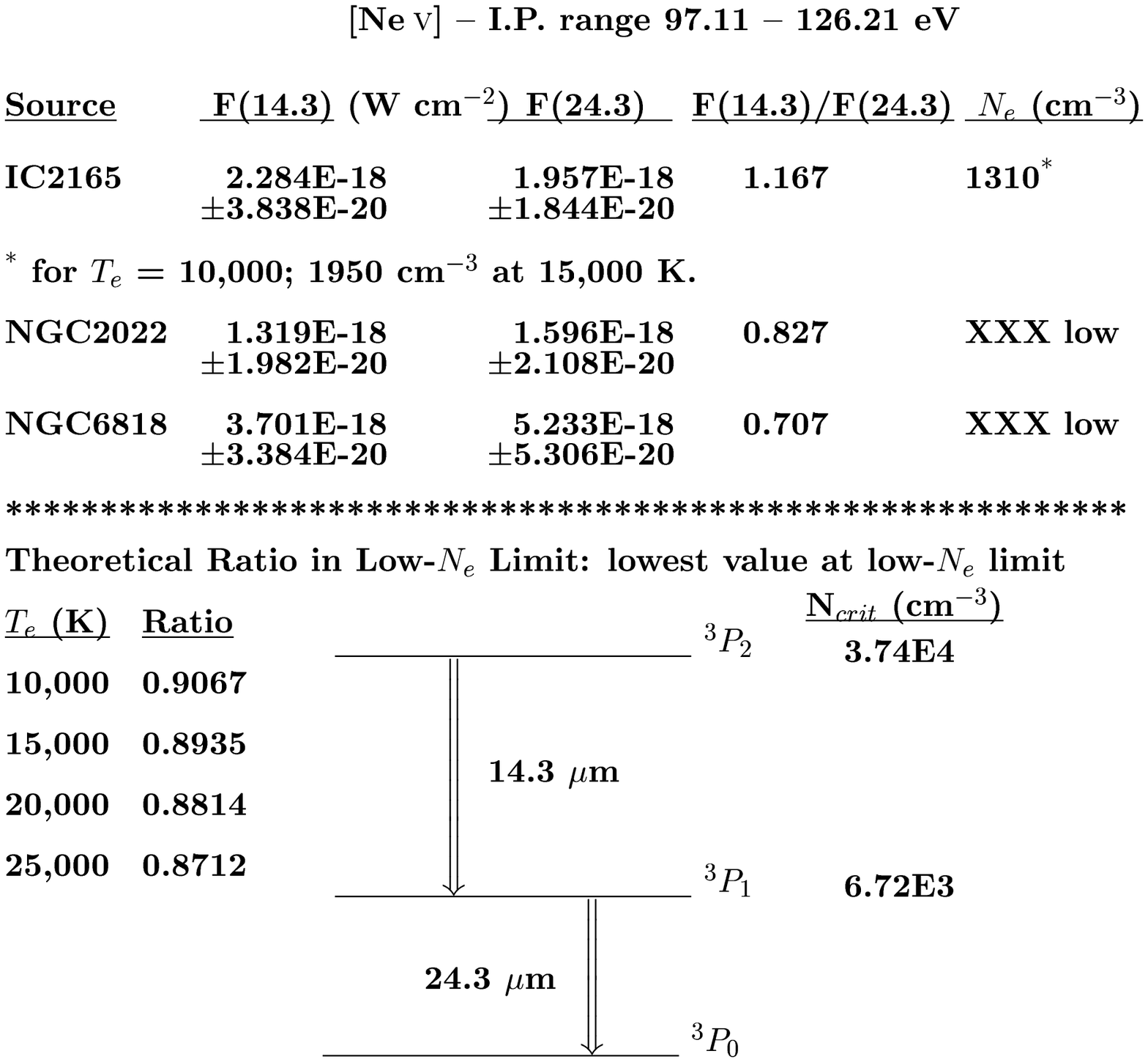}{3.0truein}{0}{40}{40}{-120}{-40}
\vskip-0.6truein
\end{table}
\vskip-0.25truein
\vspace{-5pt}

\subsection{\ion{Ne}{v}}

Observations were made of the
[\ion{Ne}{v}] 14.3 and 24.3~$\mu$m lines in IC~2165, NGC~2022, and NGC~6818
(see Figs.\ 3 \& 4).
Table~6 has the measured fluxes for these data as well as the theoretical
F(14.3)/F(24.3) ratio in the low-$N_e$ limit, where the lowest value
of the ratio obtains.
For NGC~6818, the observed ratio is well below
the low-$N_e$ theoretical limit for any reasonable $T_e$;
for NGC~2022, the observed ratio is also out of bounds.
For IC~2165, we derive $N_e$~= 1310 (if $T_e$~= 10,000~K)
and 1950~cm$^{-3}$ (if $T_e$~= 15,000~K) from Fig.~5
(see Table~6).
Compared to $N_e$-values derived for IC~2165 by Stanghellini \& Kaler
(1989),
albeit for other ions, the values here are very low.

We are aware of three other measurements of
$N_e$ from the [\ion{Ne}{v}] ratio using ISO data.
For the Seyfert galaxy NGC~4151, Alexander et~al.\ (1999) found an
observed ratio that is close to the low-$N_e$ limit
For the PNs NGC~7027 and NGC~6302, van Hoof et~al.\ (2000), found
$N_e$ of 26,900 and 12,300~cm$^{-3}$  using $T_e$ $\sim$20,000~K.
Compared to $N_e$-values derived from optical data for these PNs
for other ions (Stanghellini \& Kaler 1989),
the IR values are lower.
The fact that these previously published [\ion{Ne}{v}] ratios and
our new IC~2165 result yield generally lower $N_e$-values
than other methods tends to lend support to
our conclusion that the
low-$N_e$ theoretical
limit may need re-examination per our XXX~low results.
Just because the NGC~7027, NGC~6302, and IC~2165 data lead to
``legal, in-theoretical-bounds values"
does not mean the inferred $N_e$s are correct and does not
rule out the possibility that the collision strengths between the $^3P$
levels
may need revision.

\subsection{\ion{Mg}{v}}

Observations were made of the
[\ion{Mg}{v}] 5.61 and 13.5~$\mu$m lines in IC~2165.
Table~7 presents the measured fluxes as well as the theoretical
F(13.5)/F(5.61) ratio in the low-$N_e$ limit, where the highest value
of the ratio obtains.
Again, the observed ratio is out of the theoretical bounds.

\vspace{-5pt}
\begin{table}
\begin{center}
    \caption{~~~~~}
\plotfiddle{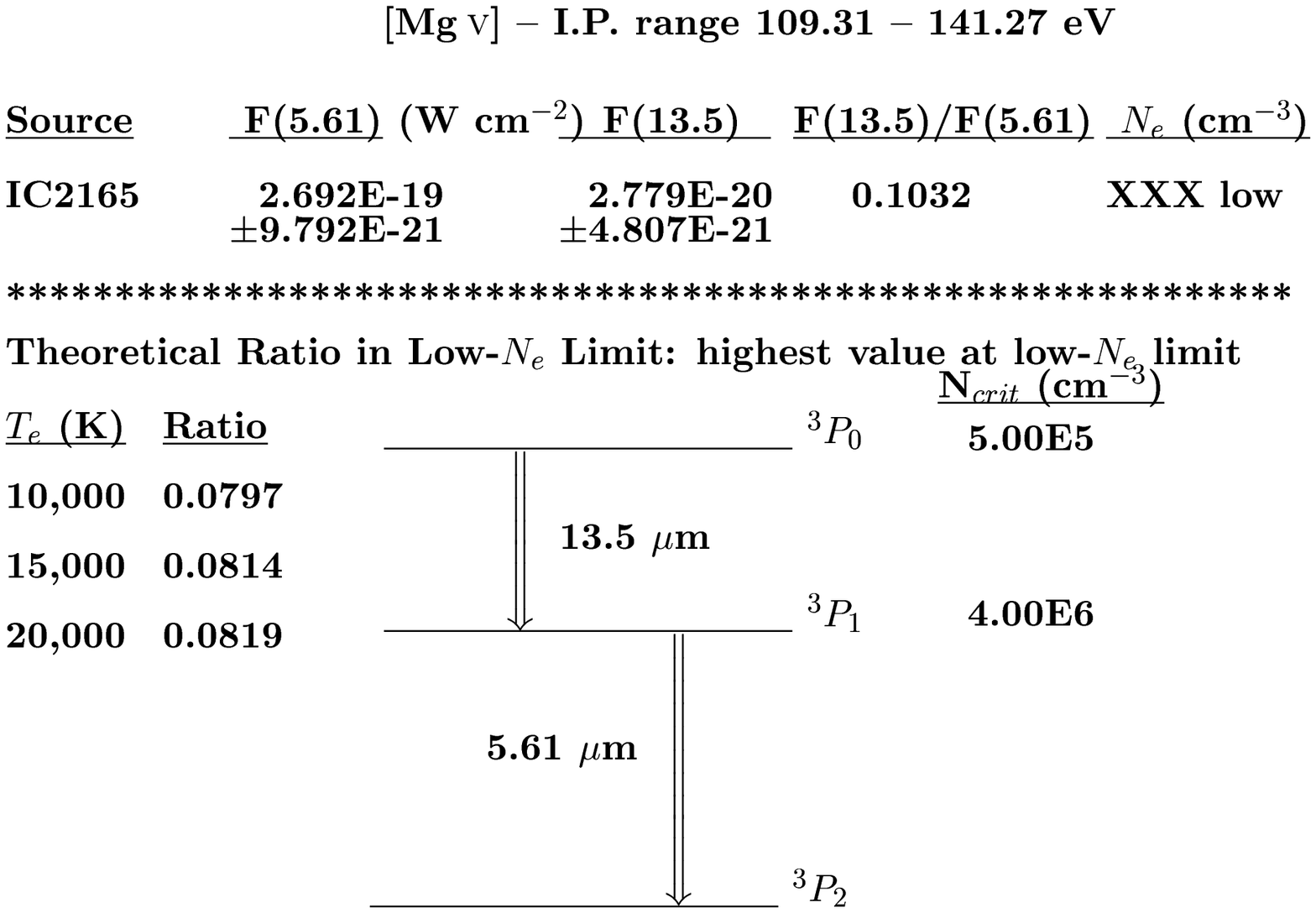}{3.0truein}{0}{40}{40}{-120}{-40}
\end{center}
\vskip-1.5truein
\end{table}

\vspace{-15pt}
\section{Observations of a Strong Unidentified IR Emission Line}

We have detected an
unidentified (uid) strong emission line in an ISO  SWS02 spectrum of the
Orion Nebula that was obtained for our Guest Investigator program
on 1997 October 12 (UT),  see Figure~6.
The 14$''$$\times$20$''$ aperture was centered at
$\alpha$, $\delta$ = $05^{\rm h}35^{\rm m}14\fs71$,
--05$^{\rm o}$23\arcmin41\farcs5)
(equinox J2000).
This position is the center of position 1SW,
18.5$''$ S and 26.2$''$ W of
$\theta$$^1$Ori~C,
observed with HST (see Rubin et~al.\ 1997).
The orientation
of the long axis of the aperture was at PA
190.28$^{\rm o}$,
determined  by the time of the actual observation.
With a correction for V$_{helio}$~= 17~km~s$^{-1}$ from
emission lines arising in the main ionization zone (O'Dell et~al.\ 1993),
the line has a rest wavelength
2.89350$\pm$0.00003~$\mu$m (3456.02 cm$^{-1}$); it
is a factor of 3.6 weaker than the nearby,
\ion{H}{i} 11--5 line (2.8728~$\mu$m).
The average surface brightness in this aperture was
3.9E-18 W~m$^{-2}$arcsec$^{-2}$.


\begin{figure}
\plotfiddle{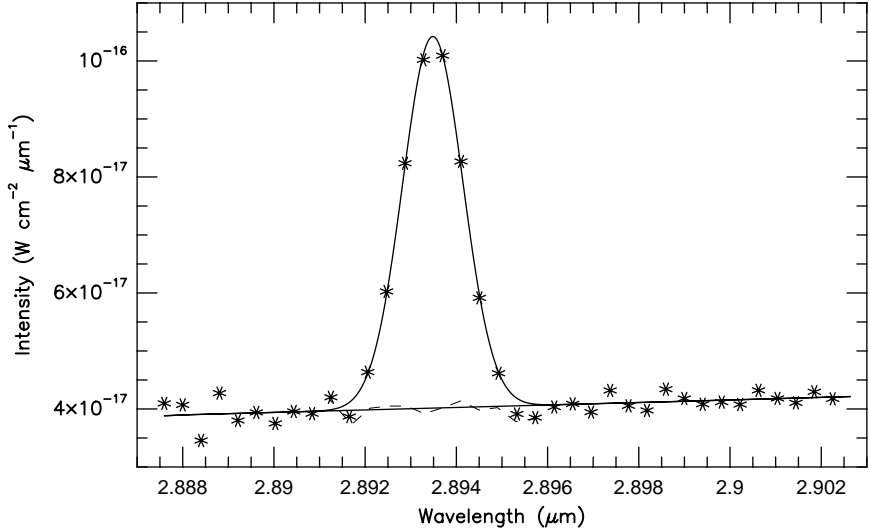}{3.5truein}{90}{50}{50}{200}{10}
\vskip-0.9truein
\caption{ISO SWS02 spectrum of the 2.8935~$\mu$m line in the Orion Nebula
taken at position 1SW (see text for details).
Observations were made with a 14$''$$\times$20$''$ aperture.
The data (asterisks) have been fit with a Gaussian and a linear baseline.
The residuals, data point minus curve fit, are indicated as the dashed
line.}
\end{figure}


\begin{figure}
\plotfiddle{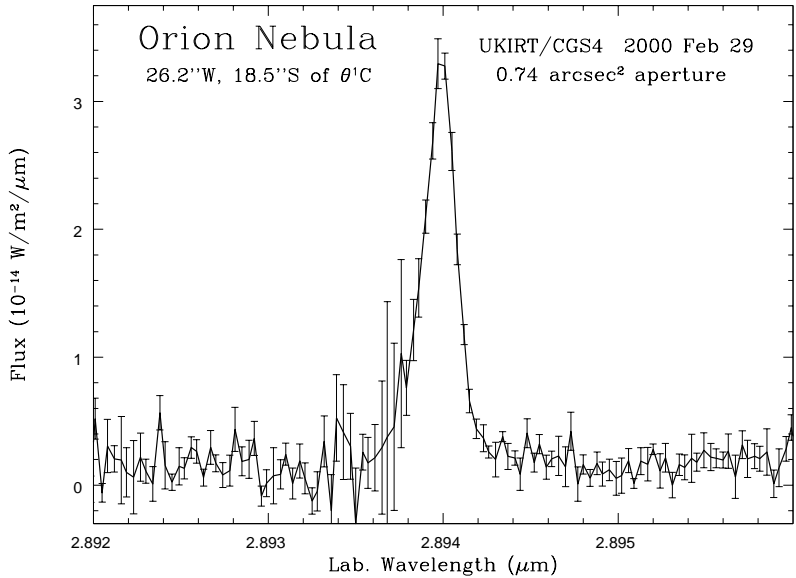}{3.5truein}{270}{45}{45}{-170}{300}
\vskip-0.5truein
\vspace{-15pt}
\caption{Portion of UKIRT long-slit spectrum of the 2.8935~$\mu$m line in
the Orion Nebula that is centered at position 1SW (see text for details).
This 1-D plot represents the spectrum through a 0.74~arcsec$^2$
aperture.
Error bars are shown.}
\vskip-0.2truein
\end{figure}

We do not detect the uid line in our SWS02 spectra of any of the
five bright PNs, which we observed for a comparable time.
These are the PNs discussed earlier plus NGC~7009 (the Saturn Nebula).
We have also examined many of the ISO archival spectra taken by
others of the Orion Nebula as well as several other PNs.
We find no indication of the uid line in the PNs examined, which included
NGC 7027,  NGC 6543, the low ionization PNs IC 418 and BD +30~3639,
as well as the bright \ion{H}{ii} region M~17.
Unlike our 1SW position (well in the heart of the ionized region),
most of the other ISO Orion spectra were taken at positions where molecular/
photodissociation (PDR)/neutral species would be expected to dominate.
There is some indication of the uid line in the following two archival
spectra:
\vskip0.05truein

\noindent
$\bullet$ an SWS01 low-resolution scan of ``Orion Bar d2" (PI Drapatz)
$\alpha$, $\delta$ =\break
$05^{\rm h}35^{\rm m}21\fs40$,
--05$^{\rm o}$25\arcmin40\farcs1 (J2000),
taken 1997 October 11.
The exposure was 1912 sec.\ with a 20$''$$\times$33$''$ aperture.
The long axis of the aperture was at PA~= 189.87$^{\rm o}$.
We measure a line flux of 3.25E-19$\pm$0.75E-19 W~cm$^{-2}$.

\vskip0.05truein

\noindent
$\bullet$ an SWS06 scan of ``Orion-BN" (PI Cernicharo)
$\alpha$, $\delta$ = $05^{\rm h}35^{\rm m}14\fs20$,\break
--05$^{\rm o}$22\arcmin23\farcs6,
taken 1997 October 12 (UT).
The exposure was 7598 sec.\ with a 20$''$$\times$27$''$ aperture.
The long axis of the aperture was at PA~= 190.36$^{\rm o}$.
We measure a line flux of 4.11E-20$\pm$1.22E-20 W~cm$^{-2}$.
\vskip0.05truein

Spectra of Orion taken at the United Kingdom Infrared Telescope
(UKIRT) confirm the presence of this strong uid line.  A long-slit
observation (80$''$$\times$1$''$)
centered at 1SW and pointing through $\theta$$^1$Ori~C
shows that the emission is spatially extended and appears to be coincident
with the brightest part of the \ion{H}{ii} region.
The highest surface brightness occurs at 1SW:
1.2E-17 W~m$^{-2}$arcsec$^{-2}$
($\pm$20\%) with FWHM~= 24 km~s$^{-1}$ (resolved).
Figure~7 shows a spectrum at this location for which
the wavelength scale has not been corrected for the Doppler velocity
of the source.  When the correction is made, the
rest wavelength is 2.8935~$\mu$m.
A second observation was made with the slit passing through the
proplyd P159-350 (O'Dell \& Wen 1994) and pointing toward $\theta$$^1$Ori~C.
This spectrum also shows extended emission in the uid line with
no significant enhancement of the line at the proplyd location.
Because P159-350 has $N_e$ at least as high as 10$^5$~cm$^{-3}$ as measured
by the [\ion{C}{iii}] 1909/1907 flux ratio from GO-7514 HST STIS program
(PI Rubin) (Henney \& O'Dell 1999),
it is a likely indication that the uid line emissivity is not
strongly enhanced at higher electron density, which  may indicate
the line is collisionally excited and not a recombination line.

We checked against lists of wavelengths of atomic ions and molecules.
For this, the atomic line list provided at www.pa.uky.edu/$\sim$peter/atomic
by Peter van Hoof was a particularly useful resource.
There were two ionic lines that were very close to the uid wavelength and
that were ``reasonable" candidates.  The first (best) candidate is:
[\ion{Cr}{iii}]   2.89349$\pm$0.00020~$\mu$m
($3d^4$~$^3G_4$ $\rightarrow$ $3d^4$~$^3H_5$) arising from
20851.87~cm$^{-1}$ ($\sim$2.6~eV) above ground (Ekberg 1997).
Chromium has a solar number abundance log [Cr/H] + 12~= 5.67
(Grevesse \& Sauval 1998).
The I.P.\ range for Cr$^{++}$ is 16.49~-- 30.96~eV, which would make
Cr$^{++}$
one of the dominant Cr species in Orion although the fraction of
Cr$^{+3}$ may be higher.
However, we do not believe that this
[\ion{Cr}{iii}] line is a likely identification
because there are other [\ion{Cr}{iii}]
lines for which we have upper limits.
These are [\ion{Cr}{iii}]   2.77830$\pm$0.00020~$\mu$m
($3d^4~^3G_5$ $\rightarrow$ $3d^4~^3H_5$) arising
20995.16~cm$^{-1}$ above ground;
2.88560$\pm$0.00022~$\mu$m
($3d^4~^3G_5$ $\rightarrow$ $3d^4~^3H_6$); and
2.91564$\pm$0.00020~$\mu$m
($3d^4~^3G_3$ $\rightarrow$ $3d^4~^3H_4$) arising
20702.45~cm$^{-1}$ above ground.
There was no hint of these lines in various SWS02 bandpasses that we
observed covering these 3 other transitions.  From the similarity of
the levels within the multiplet and $\Delta$J of 0 or -1, it is difficult
to understand how the 2.89349 line would be so strong and all the others
absent.

The second close coincidence is an [\ion{Fe}{v}]  line at
2.8934$\pm$0.00075~$\mu$m
($3d^4~^3G_5$ $\rightarrow$ $3d^4~^3F_4$) arising from
30430.10~cm$^{-1}$ above ground.
\ion{Fe}{v} is isoelectronic with \ion{Cr}{iii}.
The I.P.\ range for Fe$^{+4}$ is 54.8~-- 75.0~eV, which
is higher than the He$^{++}$ limit of 54.4~eV.
There is a stringent upper limit on the ionizing
flux greater than 54.4~eV in Orion by virtue of deep spectroscopy
that set a ``record" limit on the flux ratio \ion{He}{ii} 4686/H$\beta$
$<$ 7$\times$10$^{-5}$
(Baldwin et~al.\ 2000).
Thus, there should be negligible \ion{Fe}{v} in Orion.

Our assessment of what ionization range might produce the uid line
is that it is likely from a species having a production I.P.\ between
13.6 and 54.4~eV because it is not seen in the PNs we examined.
However the uid line is also not detected in the \ion{H}{ii} region M~17 or
the low-ionization PNs  IC~418 and BD~+30~3639.
In Orion, the uid line is much stronger in the ionized region and
only marginally present (if present at all) in the PDR/molecular/neutral
zones.
The fact that it is found in only the Orion Nebula reminds one of
the uniqueness of proplyds to Orion (at least before they were
found in some other nebulae).

There is a reward offered for the correct identification of this
uid line.  The details were presented at the meeting and will be provided
to anyone interested by writing to the first author.

\section{Upcoming and Planned IR Missions}

In this section, we discuss planned
IR spectroscopic capability
with a major emphasis on NASA and ESA projects.  There are some comparisons
with the past projects/missions such as KAO, IRAS, and ISO.
With one exception,
we do not cover ground-based observatories but do remind the reader of
excellent spectroscopic capabilities to be afforded by these facilities
(e.g.,
Gemini North/South, Keck, and the VLTs).

\subsection{Space Infrared Telescope Facility (SIRTF)}

This is a NASA space observatory with an 0.85-m telescope.
It will have three cryogenically-cooled instruments.
The launch date is expected in
July 2002
with a goal of a lifetime in excess of 5 years.
SIRTF will be in an Earth-trailing, heliocentric orbit.
The main spectroscopic instrument is the Infrared Spectrograph (IRS)
(PI Jim Houck).  There will be 4 separate modules that cover 5.3~--
40~$\mu$m at resolution (R) from 60~-- 120
(low-R mode)
and a higher resolution (R~= 600) capability from 10~-- 37~$\mu$m.
It will provide a major advance in photometric sensitivity as can be seen
in Figure~8 which shows comparisons with KAO, IRAS, ISO, and SOFIA.
The angular resolution will be an improvement over ISO
but somewhat less
than the KAO
as is seen in Figure~9.
The URL http://sirtf.caltech.edu
provides further \pagebreak details.

\enlargethispage{20pt}

\subsection{Stratospheric Observatory for Infrared Astronomy (SOFIA)}
This is a joint project of NASA and the German aerospace center DLR.
SOFIA
will be the successor to the KAO.
The facility is a modified Boeing  747SP airplane with a 2.5-meter
telescope.
First light is expected in late 2004, with general observing
commencing in 2005.
The planned lifetime is 20 years and the home base is
Moffett Field at the NASA/Ames Research Center.

The photometric sensitivity is shown
in Figure~8 in comparison  with other missions.
The angular resolution will be a substantial improvement, especially if
the goal (solid curve) is attained (see Figure~9).
The complement of First Light Instruments now being built
are all listed below, but we describe here only those with
medium to high spectral resolution capability.
Figure~10 shows the instrument coverage, along with those of SIRTF,
in R--$\lambda$ space.

\noindent
\underbar{CAMERAS}: There are 3 facility instruments~--
HAWC (PI Harper),
FORCAST (PI Herter),
and FLITECAM (PI McLean).
FLITECAM will provide moderate resolution spectroscopy
(R~$\sim$1000~-- 2000) from 1~-- 5.5~$\mu$m.

\noindent
\underbar{IMAGING PHOTOMETER}: There is one Principal Investigator (PI)
instrument, HIPO (PI Dunham).

\noindent
\underbar{SPECTROGRAPHS}: There are one facility instrument,
{\bf (1)} below, and five PI instruments {\bf (2)--(6)} below~--

\noindent
{\bf (1)} the  Airborne Infrared Echelle Spectrometer  (AIRES)
(PI Erickson)   with
ultimate design goals from 17~-- 210~$\mu$m with R~$\sim$ 10,000
(higher at 17~$\mu$m).
\begin{figure}[h]
\plotfiddle{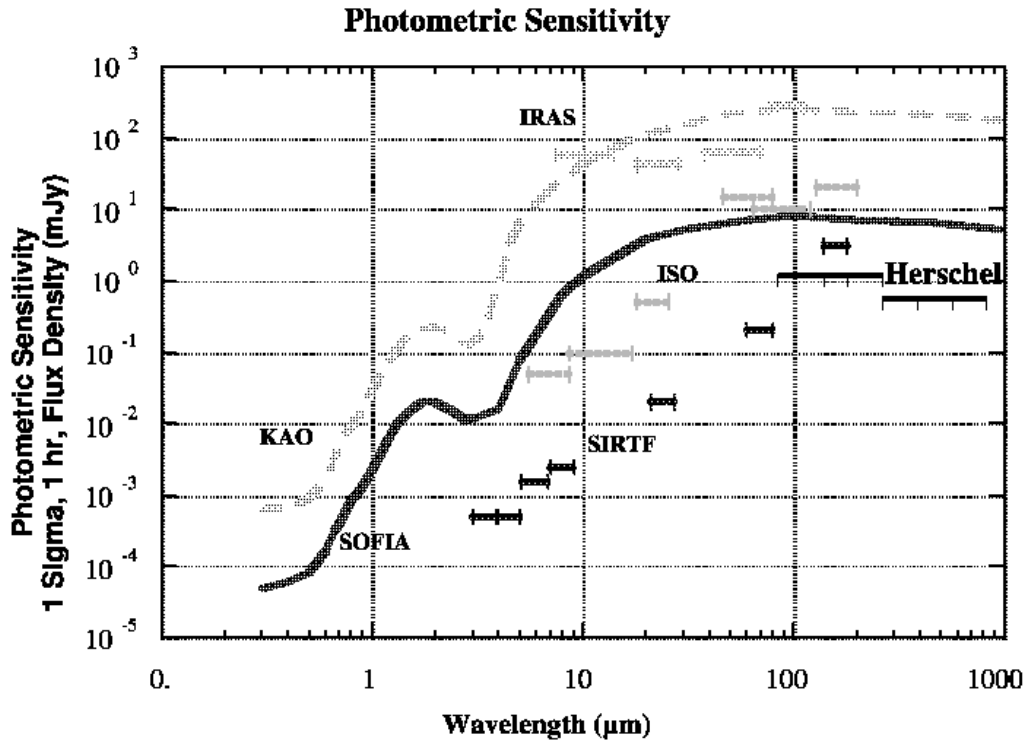}{248pt}{270}{55}{55}{-220}{300}
\vspace{20pt}
\caption{The photometric, one-$\sigma$, point source sensitivities
after one hour of integration for IRAS, KAO, ISO, SIRTF, SOFIA, and
Herschel.   Graphics from http://sofia.arc.nasa.gov.}
\vspace{-10pt}
\end{figure}

\noindent
{\bf (2)} the CAltech Submillimeter and far-Infrared MIxing Receiver
(CASIMIR) (PI  Zmuidzinas)
that is a heterodyne receiver which
will operate from 250~-- 600~$\mu$m with high spectral resolution.

\noindent
{\bf (3)} the Echelon Cross Echelle Spectrometer (EXES) (PI Lacy)~--
a grating spectrograph that will operate from 5~-- 28~$\mu$m with
high (R~$\sim$ 10$^5$), medium (R~$\sim$ 10$^4$), and
low (R~$\sim$ 3000) spectral resolution.

\noindent
{\bf (4)} the Field Imaging Far-Infrared Line Spectrometer
(FIFI LS) (PI Poglitsch).  This will cover
from 45~-- 210~$\mu$m with R~$\sim$ 1700.

\noindent
{\bf (5)} the German Receiver for Astronomy at Terahertz Frequencies
(GREAT) (PI Guesten).
This is also a heterodyne receiver and
will operate from 75~-- 250~$\mu$m with high R.

\noindent
{\bf (6)} the Submillimeter And Far InfraRed Experiment
(SAFIRE) (PI Moseley).
This is an imaging Fabry-Perot bolometer array spectrometer
which will operate from 145~-- 655~$\mu$m with an ultimate
goal of reaching R~= 10,000.

\vskip0.1truein

Further details are available via the SOFIA web site at\break
http://sofia.arc.nasa.gov/


\begin{figure}
\plotfiddle{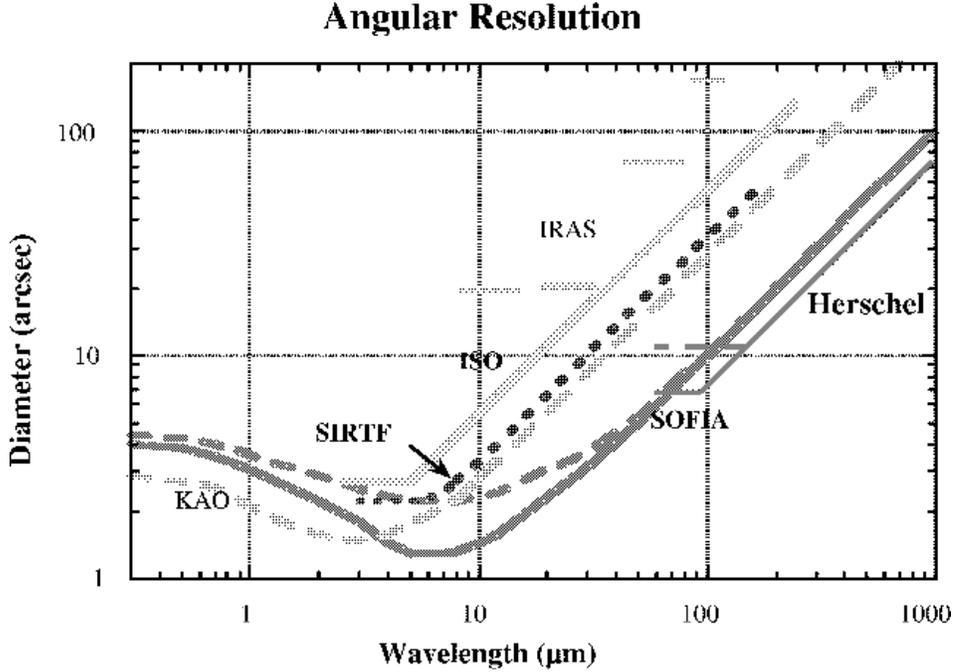}{4.0truein}{270}{55}{55}{-240}{330}
\vskip-0.5truein
\caption{The spatial resolutions ($\sim$50\% encircled energy diameter)
for IRAS, KAO, ISO, SIRTF, SOFIA, and Herschel.   Graphics from
http://sofia.arc.nasa.gov.}
\vskip-0.1truein
\end{figure}

\subsection{Herschel Space Observatory (HERSCHEL)}

Formerly called FIRST, this is an ESA ``Cornerstone Mission"
with substantial NASA participation.
It consists of a 3.5-m diameter telescope
radiatively cooled to $\sim$80~K with coverage
in the 60~-- 670~$\mu$m range.
After launch, expected to be in 2007, it will be positioned to orbit
around the Earth-Sun L2 Lagrangian point (located 1.5$\times$10$^6$~km in
the
anti-Sun direction from Earth).
It is planned to operate for at least three years.
There will be three instruments in the
helium cooled focal plane:
{\bf (1)} SPIRE, a long-wavelength (200~-- 670~$\mu$m) camera and
low-resolution
spectrometer;
{\bf (2)}
PACS, a short-wavelength (60~-- 210~$\mu$m) camera and low-resolution
spectrometer (R$\sim$1700); and
{\bf (3)}
HIFI, a high-resolution (heterodyne) spectrometer
(R similar to CASIMIR),
that will have contiguous coverage from 240~-- 625~$\mu$m
(480~-- 1250 GHz) plus a channel from 157~-- 213~$\mu$m (1410~--1910 GHz).
NASA is contributing detectors and components for two of the instruments
and plans to support US science participation through a US Herschel
Science Center at the Infrared Processing and Analysis Center (IPAC).
For more detailed information see
http://sci.esa.int/home/herschel/


\begin{figure}
\plotfiddle{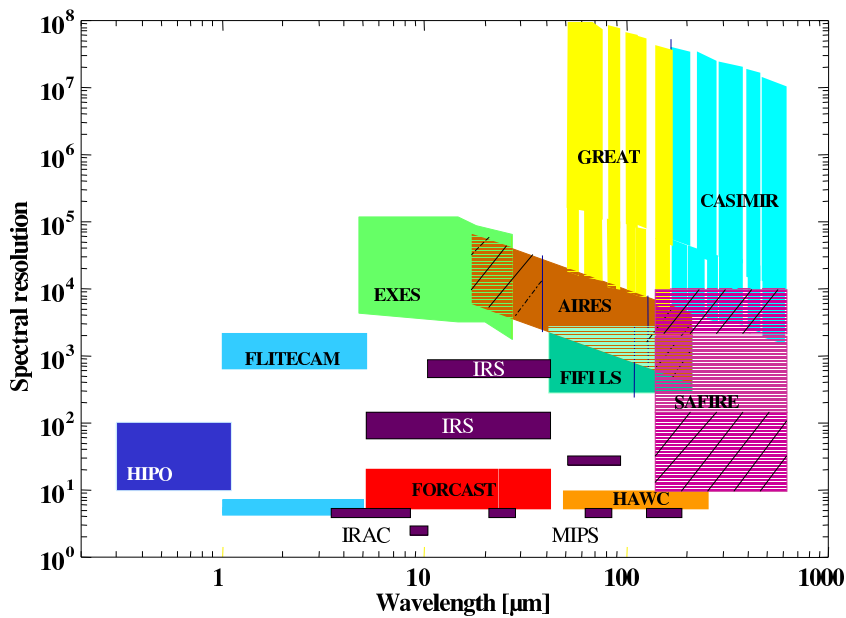}{4.0truein}{270}{55}{55}{-240}{350}
\vskip-.7truein
\caption{The spectral coverage and resolution that will be possible
on SOFIA (first-light instruments only) and SIRTF. The science
instruments shown are those for SOFIA, except for IRS, IRAC, and
MIPS. Graphics from http://sofia.arc.nasa.gov.}
\vskip-0.1truein
\end{figure}

\subsection{Planned IR Missions and the US Decadal Report 2001--2010}

Recently the latest US Astronomy Decadal Report was published
entitled ``Astronomy and Astrophysics in the New Millennium"
(National Research Council, NRC 2001).
This often is referred to as
the McKee/Taylor Report
after the co-chairs of the Astronomy and Astrophysics Survey Committee.
The Report can be accessed on-line at
www.nap.edu/books/0309070317/html.
Among other things, the report prioritized equipment initiatives
(combining ground and space).  Before continuing, we note that the
committee reaffirmed the \linebreak recommendations of the prior
Survey Committee (NRC, 1991) and endorsed the completion of SIRTF and SOFIA
as well as other projects (NRC, 2001, p.\ 24).

The highest priority of the Major new initiatives is the
Next Generation Space Telescope (NGST) (NRC 2001, p.\ 37).
Below we briefly discuss NGST, as well as two other recommendations
with major IR spectroscopic promise~-- the
\#2 and \#7 ranked  Major initiatives,
the Giant Segmented Mirror Telescope (GSMT) and
the Single Aperture Far Infrared (SAFIR) Observatory, respectively.

\noindent
{\bf NGST} will be a segmented, filled-aperture 6-m class telescope with
a large sunscreen.
It will be in an orbit $\ge$ 10$^6$~km from Earth to achieve very cold
operating temperatures.
The goal is for NGST to cover the range $\sim$ 0.6~-- 27~$\mu$m.
It is expected to be a factor of
10$^4$~-- 100 more sensitive than SIRTF (from 4~-- 20~$\mu$m).
Spectral resolution is expected to be as high as 5000.

\noindent
{\bf GSMT} will be a ground-based, 30-m class telescope.
Through adaptive optics, it will have diffraction-limited imaging
in atmospheric windows from 1~-- 25~$\mu$m.
The full wavelength goal is 0.3~-- 25~$\mu$m.
Spectral resolution is expected to be as high as R~= 10$^5$.
Unlike the space observatories planned, it will have the capability of
adding
new instruments.

\noindent
{\bf SAFIR} will be an 8-m class, passively cooled space telescope
that is planning to cover the spectrum from 30~-- 300~$\mu$m.
It is anticipated to provide 2 to 10 times the spatial resolution
of SIRTF, SOFIA, or Herschel.
Even more dramatic will be SAFIR's increase in sensitivity compared
with these upcoming missions, reaching several orders of magnitude
depending on wavelength  (see Fig.~3.2 on p.\ 103 of NRC 2001).
Spectral resolution is expected to be as high as  1000.

\section{Summary and Conclusions}

With regard to our efforts to derive electron densities,
there are several instances of
the observed line flux ratio being clearly out of range of the theoretical
predictions using current atomic data.
This is most dramatically illustrated with the
[\ion{Ne}{v}] (14.3/24.3) flux ratio for NGC~6818.
These are both strong lines observed with signal-to-noise ratios better than 100.
The observed flux ratio is 0.707,
which is significantly less than
the predicted ratio using collision strengths from
Lennon \& Burke (1994).
Because the line ratio near the low-$N_e$ limit depends on collision
strengths and \underbar{not} the transition probabilities (A-values),
this result may point toward a need to reexamine the collision strengths.

A very rapid benefit of the type of interaction the Conference
was to promote occurred in this case.
Two participants, Nigel Badnell and Keith Berrington,
had made new computations for the \ion{Ne}{v} effective
collision strengths, including the relevant transitions.
The former work had just been published (Griffin \& Badnell 2000).
Their paper provides the most detailed calculations to date.
Their recommended values for each transition are
the last entries in their Table 5.
In the low-$N_e$ limit, F(14.3)/F(24.3) should scale as
(C$_{13}$/(C$_{12}$~+ C$_{13}$))$\times$(24.3175/14.3217),
where
C$_{12}$~= $\Upsilon$($^3P_0-^3P_1$)~$exp(-\chi_{12}/kT_e)$
and
C$_{13}$~= $\Upsilon$($^3P_0-^3P_2$)~$exp(-\chi_{13}/kT_e)$.
The $\chi$s are the energy level differences and $k$ is
the Boltzmann constant.
For $T_e$~= 10,000~K, this gives a flux ratio 0.8090,
while our detailed solution for the 5 lowest levels yields 0.8097.
Although not material for this calculation,
we did include
$\Upsilon$($^1D_2-^1S_0$)
(transition between lowest levels 4 and 5)
that had been omitted in their publication.
The missing entry is 0.545 at 10,000~K (Badnell, private communication).
We use the most accurate wavelengths for the two [\ion{Ne}{v}] IR lines
measured from high spectral resolution ISO observations
of a PN (Feuchtgruber et~al.\ 1997) to provide energy levels 2 and 3
in our code.
Their paper is a fine example of how astrophysical measurements
can actually do better than any laboratory measurement or quantal
calculations to determine fundamental atomic data.

The result of using the Griffin \& Badnell (2000) $\Upsilon$s
is gratifying,
bringing closer agreement with the observations (see Table 6)
although the NGC~6818 observed ratio is still
smaller than the low-$N_e$ limit for $T_e$ 10,000~K or any reasonable
$T_e$ (using adjacent 6,300 and 25,100~K entries in their table~5).

We use CS's for \ion{S}{iii}, \ion{Ar}{iii}, and \ion{Ar}{v} from
Galavis~et~al.\ (1995); for \ion{Ne}{iii} and
\ion{Mg}{v} from Butler \& Zeippen (1994).
As a result of this Conference, we learned of the new
calculations of \ion{Ne}{iii}  CS's (McLaughlin \& Bell 2000).
When we use these instead of Butler \& Zeippen values, we find that the
F(36.0)/F(15.5) ratio in the low-$N_e$ limit,
for $T_e$~= 10,000~K, is increased slightly to 0.090.
This makes the discrepancy with the observed ratios only slightly less
(see Table 4).

Recently Dinerstein (2001) identified
two long-standing cases of IR uid lines~--
at 2.199 and 2.287~$\mu$m.
These were seen in the spectra of PNs (e.g.,
Geballe, Burton, \& Isaacman 1991 and references therein).
She identified the former as
[\ion{Kr}{iii}] $^3P_1$--$^3P_2$,
and the latter as
[\ion{Se}{iv}] $^2P_{3/2}$--$^2P_{1/2}$.
The identification of such heavy elements in the IR
suggests that there is much future work to be done in
the Conference ``fields of endeavor".
To extract astrophysical knowledge, e.g., nebular elemental abundances
of such species, it will be necessary to observe/identify spectral lines,
add to the atomic database for many more heavy elements,
and add their treatment to photoionization codes.
In addition, there can be expanded $N_e$ diagnostic capability
by enlarging the number of ionic species observed that have $^3P$ ground
states.

As this work shows, progress in the future will also depend very much
on the ability to obtain co-spatial observations of the line pairs
in the IR used as an $N_e$ diagnostic.
Thus it is important that future astronomical facilities be able to
do this and maintain a level of flux calibration sufficient
for analyses of electron density, such as we had hoped to perform.

\acknowledgements
This research was supported by NASA through data analysis grants
to the ISO General Observer program and
by AURA/STScI grant related to GO-6792.
We thank Robin Ellis for help with the ISO data processing
and Jackie Davidson for assistance with Figures~8--10.
We made use of Richard Shaw's program at
http://ra.stsci.edu/nebular/ionic.html
for several of the ions.
The atomic line list at www.pa.uky.edu/$\sim$peter/atomic
by Peter van Hoof was also used.
RHR acknowledges NASA/Ames Research Center cooperative agreement
NCC2-9018 with Orion Enterprises
and thanks Scott McNealy for providing a Sun workstation.

\end{document}